# Black holes of intermediate masses in globular clusters: constraints on a spin of a black hole.

Buliga S.D.[1], Globina V.I.[2], Gnedin Yu.N.[1,2], Natsvlishvili T.M.[1], Piotrovich M.Yu.[1], Shaht N.A.[1]

(1) Central astronomical observatory at Pulkovo, St.-Petersburg, Russia.
(2) St.-Petersburg State Polytechnic University, St.-Petersburg, Russia.

**Abstract**

In this paper we determined values of a spin of central black holes of the intermediate masses in globular clusters. For determination of value of a spin we used the known relation between the kinetic power of the relativistic jet and observable radio-luminosity of the region near to a central black hole, and our estimates have based on the known Blandford-Znajek mechanism. The value of a magnetic field strength near the event horizon of a black hole was derived via magnetic coupling mechanism. Accretion rate $\dot{M}$ was derived using Bondi-Hoyle mechanism.

**Keywords:** black holes, magnetic field, globular clusters.

## 1. Introduction

One of central problems of the modern astronomy is search of black holes of intermediate masses ($10 M_\odot < M_{BH} < 10^5 M_\odot$) [1,2,4,5].

The central objects of globular clusters are considered now as the most popular candidates for a class of such objects.

The well-known relation between mass of central object and stellar velocity dispersion allows to determine masses of these objects with quite enough accuracy. For typical massive globular cluster having a dispersion of the order of 10 km/s, a black hole of about 2x10^[3] M{Sun} is expected. Detection of black holes of the intermediate mass will allow to fill in a gap existing now between black holes of stellar masses and supermassive black holes.

Relation between mass of central object and a dispersion of velocities of nearby stars represent the good tool for definition of mass of this object. It is known, at least some globular clusters with central nuclei which are confidently identified with black holes of intermediate masses and well correspond to the mass-dispersion of velocities dependence [6, 7] determined for supermassive black holes.

Recently Lu and Kong [8] have presented the results of the deep radio survey of two galactic globular clusters 47 Tuc and $\omega$ Cen. Observations have been done via Australia Telescope Compact Array (ATCA). Authors have discovered concentration of a radio flux to centres of globular clusters at the level of $2.5\sigma$. For 47 Tuc and $\omega$ Cen such flux (at the level $3\sigma$) was $40\mu J$ and $20\mu J$ correspondingly. Using well-known fundamental relations between the radio and X-ray luminosities and the mass of a supermassive black hole [9], authors of [8] have derived values of masses of central objects in the given globular clusters. The obtained values correspond well to masses of black holes at the level of $(10^3 \div 10^4) M_\odot$. In this paper similar data for other globular clusters (Tab.1 from [8]) are also presented.

The purpose of this paper is to determine the constraints on value of the dimensionless specific angular momentum (spin) $a_*$ of black holes of intermediate masses in globular clusters. Let us remind, that in the standard system of units $G = \hbar = c = 1$ parameter $a_* = 1$ for extremely rotating Kerr black hole and $a_* = 0$ for Schwarzschild black hole. We use the following expression for the spin $a_*$ obtained in [10]:

$$a_* = \eta \left(\frac{L_j}{10^{44}}\right)^{0.5} \left(\frac{10^4}{B_H}\right)\left(\frac{10^8 M_\odot}{M_{BH}}\right) \qquad (1)$$

where - $L_j$ is the energy power of a jet, $B_H$ - value of a magnetic field strength in ergosphere of a black hole, $M_{BH}$ - mass of a black hole. The coefficient $\eta$ is modelling parameter which is equal to $\eta = \sqrt{5}$ for the physical model of jet developed by Blandford and Znajek [11]. For the hybrid model uniting Blandford -Znajek and Blandford-Payne mechanisms [12], value $\eta = (1.05)^{1/2}$.

The key point in determination of a spin of a black hole is a value of a magnetic field strength $B_H$, which is responsible for generation of relativistic jet. Such value can be estimated via the magnetic coupling (MC) model, developed in [13, 14, 15, 16]. In this model the magnetic field on event horizon of a black hole originates from interaction between accreting matter and a rotating black hole. As a result value of such field is derived from a relation between accretion rate $\dot{M}$, mass of a black hole $M_{BH}$ and its spin $a_*$:

$$B_H = \frac{\sqrt{2k\dot{M}c}}{R_H}; \quad R_H = \frac{GM_{BH}}{c^2}\left[1+\sqrt{1-a_*^2}\right] \qquad (2)$$

According to [8], Bondi-Hoyle accretion rate is derived as follows:

$$\dot{M} = 3.2 \times 10^{17} \left(\frac{M_{BH}}{2\times 10^3 M_\odot}\right)^2 \left(\frac{n}{0.2 cm^{-3}}\right)\left(\frac{T}{10^4 K}\right)^{-1.5} g/s \qquad (3)$$

where $n$ is the gas density in the center of a globular cluster, and $T$ is the temperature of this gas.

As to the kinetic power of the relativistic gas, it can be estimated by means of the known relation between power of jet $L_j$ and the luminosity of the central object of a globular cluster in radio-frequency range $L_R$ [19]:

$$L_j = 5.8 \times 10^{43} \left(\frac{L_R}{10^{40}}\right)^{0.4} erg/s \qquad (4)$$

Further, using the relation (1) it is possible to estimate value of a spin of a black hole of intermediate masses in globular cluster. First of all, we will estimate value of a spin, assuming the equality of densities of magnetic and a kinetic energy near to event horizon, i.e. $k=1$.

## 2. Estimation of a spin of a central black hole in globular clusters 47 Tuc and $\omega$ Cen

Detailed radio observations of globular clusters 47 Tuc and $\omega$ Cen have been made by Australia Telescope Compact Array (ATAC) in the interval from January 21 till January 25, 2010 [8]. Observations were produced simultaneously on frequencies 5.5 and 9 GHz. The maximum values of radio fluxes of radiation of central areas of these clusters (at level $\leq 3\sigma$) are equal to $F_{5GHz} = 20\mu Jy$ for $\omega$ Cen and $F_{5GHz} = 40\mu Jy$ for 47 Tuc. The obtained data allow to estimate the kinetic power of jets of these objects assuming that they are the black holes of the intermediate masses. Such estimation is made by means of equation (4). As a result we obtained $L_j = 10^{35} erg/s$ for $\omega$ Cen and $L_j = 4.9\times 10^{35} erg/s$ for the globular cluster 47 Tuc. Values of masses of central objects in these globular clusters are obtained on the basis of dynamic observations: $M_{BH} = 1.2 \times 10^4 M_\odot$ for $\omega$Cen [17] and $M_{BH} = 1.5 \times 10^3 M_\odot$ for 47 Tuc [18]. Value of a magnetic field strength near to event horizon of a black hole of the intermediate mass is estimated by means of the equation (3). For a case of equality of densities of magnetic and kinetic energy $(k=1)$ we obtain the following values of a magnetic field strength: $B_H = \dfrac{2.2 \times 10^5}{1+\sqrt{1-a_*^2}} G$ for $\omega$ Cen and $B_H = \dfrac{5.5 \times 10^5}{1+\sqrt{1-a_*^2}} G$ for 47 Tuc. The equation (1) allows to estimate the value of the spin of a central black hole in a globular cluster. For $\omega$Cen ($k=1$) it gives $a_* = 0.07$. For model of the standard Shakura-Sunyaev disk [20], $k \approx \alpha$, where $\alpha$ is a coefficient of viscosity. For traditional value of the coefficient of viscosity $\alpha \approx 0.01$ we obtained the following value of a spin of a black hole of the intermediate mass in cluster $\omega$Cen: $a_* = 0.2$.

For 47 Tuc we obtained from (1) the following equation for determination of value of the spin of a black hole:

$$\frac{a_*^2}{\left[1+\sqrt{1-a_*^2}\right]^2} = 3.6 \times 10^{-2} \qquad (5)$$

The equation (5) is obtained assuming the equality of density of magnetic and accretion energy, i.e. $k=1$. The solution of (5) gives for a central black hole in 47 Tuc the following value: $a_* = 0.35$.

## 3. The spin of the central black hole in globular cluster G1 of M31 galaxy

K.Gebhardt et al. [21] have determined a value of mass of a black hole at centre of globular cluster G1 in M31 $M_{BH} = 1.8 \times 10^4 M_\odot$. Radio observations of G1 by J.S. Ulvestad et al. [22] via VLT, have allowed to determine value of a radio flux at 8.4 GHz $F_{8GHz} = 2 \times 10^{15} W/Hz$. As a result we obtain the following value of the kinetic power of the relativistic jet: $L_j = 10^{38.3} erg/s$. For following values of the accretion rate [22], $N_H = 0.2 cm^{-3}$, $T_e = 10^4 K$ and $k=1$ we obtain an estimation of value of a magnetic field strength near the event horizon of a black hole of the intermediate mass in globular cluster G1:

$$B_H = \frac{3.35 \times 10^5}{1+\sqrt{1-a_*^2}} G \qquad (6)$$

**Table 1.**

| Globular clusters | $M_{BH}/M_\odot$ | $a_*$ | $B_H (G)$ |
|---|---|---|---|
| $\omega$ Cen | $1.2 \times 10^4$ | $\leq 0.2$ | $1.1 \times 10^5$ |
| 47 Tuc | $1.5 \times 10^3$ | 0.35 | $2.8 \times 10^5$ |
| NGC 6388 | $5.7 \times 10^3$ | 0.2 | $1.66 \times 10^5$ |
| NGC 2808 | $2.7 \times 10^3$ | 0.3 | $2.7 \times 10^5$ |
| M 15 | $2.54 \times 10^3$ | 0.2 | $2.34 \times 10^5$ |
| M 62 | $3 \times 10^3$ | 0.2 | $1.7 \times 10^5$ |
| M 80 | $1.6 \times 10^3$ | 0.35 | $2.4 \times 10^5$ |

The corresponding equation for determination of value of a spin of a black hole is

$$\frac{a_*^2}{\left[1+\sqrt{1-a_*^2}\right]^2} = 0.28 \qquad (7)$$

The solution of (7) gives following value of a spin: $a_* = 0.8$.

Results of our calculation of values of a spin of black holes of intermediate masses in other globular clusters are presented in Table 1.

## 4. Conclusions

In this paper we determined the values of a spins of black holes of the intermediate masses in the central areas of globular clusters. Results of our calculations show that values of spins of black holes of the intermediate masses do not exceed value $a_* = 0.35$. It means, that such objects are most likely black holes with a small specific moment of rotation with the exception of a black hole at centre of globular cluster G1 which value of the dimensionless angular momentum is considerably higher and is equal to $a_* = 0.8$. This estimate is obtained with assuming the equality of densities of magnetic and accretion energy near the event horizon of a black hole.

## Acknowledgements

This research was supported by the program of Presidium of RAS "Origin and Evolution of Stars and Galaxies" № 20, the program of the Department of Physical Sciences of RAS "Extended Objects in the Universe" № 16, by the Federal Target Program "Scientific and scientific-pedagogical personnel of innovative Russia" 2009-2013 (№ 02.740.11.0246) and by the grant from President of the Russian Federation "The Basic Scientific Schools" (NSh-3645.2010.2).